\pdfoutput=1
\documentclass[11pt,twoside,a4paper,cmspaper,final,collab]{cms-tdr}
\def\svnVersion{9676ea4}\def\svnDate{2026/05/15}\def\cmsCernNoTag{CERN-EP-2025-222}\def\cmsCernDate{\today}\def\cmsMessage{Published in Physical Review Letters as \href{http://dx.doi.org/10.1103/26wx-tg6f}{\doi{10.1103/26wx-tg6f}.}}
\begin{document}\cmsNoteHeader{HIN-25-009}

\newcommand {\AonA}  {\ensuremath{\text{AA}}\xspace}
\newcommand {\PbPb}  {\ensuremath{\text{PbPb}}\xspace}
\newcommand {\pPb}  {\ensuremath{\Pp\text{Pb}}\xspace}
\newcommand {\OO}  {\ensuremath{\text{OO}}\xspace}
\newcommand {\NeNe}  {\ensuremath{\text{NeNe}}\xspace}
\newcommand {\pp}    {\ensuremath{\Pp\Pp}\xspace}
\newcommand {\pA}    {\ensuremath{\Pp\text{A}}\xspace}
\newcommand{\rootsNN}{\ensuremath{\sqrt{s_{_\mathrm{NN}}}}\xspace}
\newcommand{\deta}{\ensuremath{\Delta\eta}\xspace}
\newcommand{\dphi}{\ensuremath{\Delta\phi}\xspace}
\newcommand{\dmean}[1]{\ensuremath{\langle\langle#1\rangle\rangle}\xspace}
\newcommand{\cn}[1]{\ensuremath{c_n\{#1\}}\xspace}
\newcommand{\vn}[1]{\ensuremath{v_n\{#1\}}}
\newcommand{\etsumhf}{\ensuremath{E_{\text{T, sum}}^{\mathrm{HF}}}\xspace}
\newcommand{\vsectwo}{\ensuremath{v_{2}\{2,\abs{\deta}>2\}}\xspace}
\newcommand{\vtrdtwo}{\ensuremath{v_{3}\{2,\abs{\deta}>2\}}\xspace}
\newcommand{\vsecfour}{\ensuremath{v_{2}\{4\}}\xspace}
\newcommand{\vsec}{\ensuremath{v_{2}}\xspace}
\newcommand{\vtrd}{\ensuremath{v_{3}}\xspace}
\newcommand{\vnn}{\ensuremath{v_{n}}\xspace}
\newcommand{\zvtx}{\ensuremath{z_\text{vtx}}\xspace}

\title{Observation of long-range collective flow in OO and NeNe collisions and implications for nuclear structure studies}

\date{\today}

\abstract{The long-range collective flow of particles produced in oxygen-oxygen (\OO) and neon-neon (\NeNe) collisions is measured with the CMS detector at the CERN LHC. The data samples were collected at a center-of-mass energy per nucleon pair of 5.36\TeV, with integrated luminosities of 7\nbinv and 0.8\nbinv for \OO and \NeNe collisions, respectively. Two- and four-particle azimuthal correlations are measured over nearly five units of pseudorapidity. Significant elliptic (\vsec) and triangular (\vtrd) flow harmonics are observed in both systems. The ratios of \vnn coefficients between \NeNe and \OO collisions suggest sensitivity to the intrinsic nuclear structure of the respective Ne and O ions. Hydrodynamic models with \textit{ab initio} nuclear structure inputs qualitatively reproduce the collision-overlap dependence of both the \vnn values and the \NeNe to \OO ratios. These measurements provide new constraints when modeling the interplay of nuclear structure and collective dynamics in collisions of $^{16}$O and $^{20}$Ne ions, respectively.
}

\hypersetup{
pdfauthor={CMS Collaboration},%
pdftitle={Observation of long-range collective flow in OO and NeNe collisions and implications for nuclear structure studies},
pdfsubject={CMS},
pdfkeywords={CMS, heavy ion, collective flow, small systems}}

\maketitle

Quantum chromodynamics predicts that nuclear matter undergoes a phase transition to a deconfined state of quarks and gluons, known as the quark-gluon plasma (QGP), 
at extreme temperatures and densities~\cite{Shuryak:1977ut,Collins:1974ky}. 
One of the most important probes to investigate the formation and properties of the QGP is the study of collective flow, 
which is characterized by strong correlations in the azimuthal ($\phi$) distributions of particles produced 
over wide pseudorapidity ($\eta$) ranges in relativistic nucleus-nucleus collisions. 
These correlations are quantified through a Fourier decomposition into flow harmonics (\vnn, where $n$ is the order of the harmonic).  
Experimental \vnn results in heavy ion collisions, such as lead-lead and gold-gold collisions, at the BNL RHIC~\cite{STAR:2005ryu,STAR:2009ngv,PHOBOS:2009sau} 
and CERN LHC~\cite{Chatrchyan:2011eka,Chatrchyan:2012wg,ALICE:2010suc,ATLAS:2012at,CMS:2012zex,CMS:2013bza,CMS:2024krd} can be 
quantitatively described by hydrodynamic models of a nearly perfect fluid~\cite{Ollitrault:1992bk,Heinz:2013th,Gale:2013da}. 
These models map the initial collision geometry and its event-by-event fluctuations to the observed collective phenomena 
through the hydrodynamic response of the QGP~\cite{Alver:2010dn,Schenke:2010rr,Qiu:2011hf}.

Over the past decade, similar flow-like signals have been observed in high-multiplicity proton-proton (\pp)~\cite{Khachatryan:2010gv,Aad:2015gqa,Khachatryan:2015lva,Khachatryan:2016txc,CMS:2020qul}, 
proton-nucleus (\pA)~\cite{CMS:2012qk,alice:2012qe,Aad:2012gla,LHCb:2015coe,CMS:2015yux,CMS:2018loe,CMS:2018duw,CMS:2019wiy,CMS:2022bmk,CMS:2024krd}, 
as well as in other collisions involving a light ion with a heavy ion~\cite{PHENIX:2014fnc,PHENIX:2015idk,PHENIX:2016cfs,PHENIX:2017djs,PHENIX:2018lia,PHENIX:2021ubk,STAR:2015kak,STAR:2019zaf,STAR:2022pfn,STAR:2023wmd}. 
These observations raise the question of whether QGP droplets are produced in small collision systems~\cite{Nagle:2018nvi,Dusling:2015gta}. 
A long-standing challenge in quantitatively describing the collective phenomena in \pp and \pA collisions is the limited understanding of the initial collision geometry, 
especially the strong dependence on the proton structure~\cite{Schenke:2021mxx,STAR:2023wmd}.
In addition, \pA collisions involve an asymmetric longitudinal geometry, which further complicates the theoretical modeling~\cite{Zhao:2022ugy,Bozek:2015bna}. 
In contrast to \pA collisions, where initial conditions are dominated by event-by-event fluctuations, 
symmetric collisions of light ions provide better control of the initial collision geometry, 
since it is determined primarily by the spatial distribution of nucleons rather than the partonic structure of the proton~\cite{Giacalone:2024luz}. 
This makes symmetric light ion collisions ideal for investigating the final-state collective response in small collision systems. 

Ultrarelativistic collisions of ions also offer a unique opportunity to probe nuclear structure. 
Although nuclei with a $J^{\pi}$=$0^{+}$ ground state are spherically symmetric in the laboratory frame, 
their wave functions can be described by the superposition of intrinsic structures that are averaged over all orientations~\cite{Peter:1980}. 
In ultrarelativistic ion collisions, these correlated many-body densities provide the initial spatial distributions for the subsequent evolution of the produced medium. 
By probing the initial matter distribution, collective flow measurements provide information related to nuclear intrinsic structure arising from spatial correlations in the nuclear ground state, 
which are often described in terms of deformation effects~\cite{CMS:2019cyz,Li:2019kkh,Xu:2021uar,Zhang:2021kxj,Ryssens:2023fkv,Xu:2024bdh,STAR:2024wgy,Giacalone:2024ixe}.
The application of this technique in ultrarelativistic heavy ion collisions has been used to constrain the quadrupole deformation of the uranium nucleus~\cite{STAR:2024wgy}.

The study of oxygen-oxygen (\OO) and neon-neon (\NeNe) collisions is particularly compelling 
because the two ions have similar mass numbers, leading to the expectation of similar hydrodynamic evolution. 
Differences in their flow properties are therefore expected to primarily reflect differences in the initial collision geometry, which may be influenced by the intrinsic nuclear structure of the respective ions. 
The $^{16}$O and the $^{20}$Ne nuclei are predicted to have tetrahedral and "bowling pin"-like intrinsic shapes in modern \textit{ab initio} nuclear structure calculations, where nucleons may group into 
$\alpha$-cluster correlations~\cite{Robson:1979zz,PhysRevLett.49.1119,Bijker:2014tka,Marcos:1983msl,Robledo:2011nf,Marevic:2018crl,Lim:2018huo}, 
each with two paired neutrons and two paired protons. 
Hydrodynamic simulations with \textit{ab initio} nuclear structure inputs, including nuclear lattice effective field theory, variational Monte Carlo, 
and extended quantum molecular dynamics simulations~\cite{Epelbaum:2013paa,Pieper:2002ne,Lee:2008fa}, predict that 
the \NeNe to \OO elliptic flow (\vsec) ratio is directly sensitive to the exotic deformation of $^{20}$Ne~\cite{Giacalone:2024luz}.  
In this context, \OO and \NeNe collisions provide a means to explore how nuclear structure may affect collective flow observables. 
The extent to which nuclear structure can be inferred from heavy ion collision observables remains, however, an active topic of discussion~\cite{Dobaczewski:2025rdi,Hagino:2025vxe}. 

This Letter presents measurements of collective flow in \OO and \NeNe collisions at a center-of-mass energy per nucleon pair of $\rootsNN=5.36\TeV$ with the CMS detector at the LHC. 
The \vsec and \vtrd harmonics are extracted from long-range, two-particle correlations over a large pseudorapidity range. 
The \vsec values are also determined from four-particle correlations with a different response to event-by-event geometry fluctuations than two-particle measurements.  
The harmonic values, measured for various overlap ranges, are compared to hydrodynamic models incorporating modern understanding of nuclear structure in their initial conditions. 
Tabulated results are provided in the HEPData record for this analysis~\cite{hepdata}.
During the final preparation of this Letter, the ATLAS and ALICE Collaborations submitted collective flow results in \OO and \NeNe collisions at $\rootsNN=5.36\TeV$, 
obtained with different analysis techniques and phase-space coverage than used here~\cite{ALICE:2025luc,ATLAS:2025nnt}.

The central feature of the CMS apparatus is a superconducting solenoid of 6\unit{m}
internal diameter, providing a magnetic field of 3.8\unit{T}. Within the solenoid volume,
there are four subdetectors, including a silicon
pixel and strip tracker detector, a lead tungstate crystal electromagnetic calorimeter,
and a brass and scintillator hadron calorimeter, each composed of a
barrel and two endcap sections. Iron and quartz-fiber Cherenkov hadron forward (HF)
calorimeters, which are subdivided into $0.175\times 0.175$ ($\Delta\eta\times \Delta\phi$) towers, cover the pseudorapidity range $2.9<\abs{\eta}<5.2$.
The silicon tracker measures charged particles within the range $\abs{\eta}<3.0$. 
For charged particles with $1<\pt<10\GeV$ and $\abs{\eta}<1.4$, the track resolutions are typically 1.5\% in \pt and 
25--90 (45--150)\mum in the transverse (longitudinal) impact parameter \cite{CMS:2014pgm}. 
For the extended pseudorapidity range of charged particles with $\abs{\eta}<2.4$, the track resolutions are
typically 2.5\% in \pt and 30--200 (150--950)\mum in the transverse (longitudinal) impact parameter. 
A detailed description of the CMS detector,
together with a definition of the coordinate system used and the relevant kinematic variables,
can be found in Refs. \cite{Chatrchyan:2008zzk,CMS:2023gfb}.

The \OO and \NeNe data were collected in July 2025 and correspond to integrated luminosities of 7\nbinv and 0.8\nbinv, respectively. 
The events were triggered by requiring one or more energy deposits of at least 7\GeV in either of the HF calorimeters. 
In the correlation analysis, an additional requirement is imposed that at least two towers have an energy
deposit above 4\GeV in each of the HF calorimeters. 
The events are required to contain a single primary vertex~\cite{CMS:2014pgm} (located along the beam axis at \zvtx) that is
within 15\unit{cm} of the nominal interaction point along the beam axis and 2\unit{cm} 
in the transverse plane. The requirement of only one primary vertex is used to avoid potential effects from multiple interactions in a single bunch crossing, 
i.e., pileup. The average number of interactions in a single bunch crossing is less than 0.26 (0.13) for \OO (\NeNe) collisions. 
Events are classified by collision centrality based on the sum of the total transverse momentum (\etsumhf) deposited in the HF calorimeters \cite{CMS:2011iwn}. 
The \etsumhf distribution is fitted by a Glauber-model-based calculation~\cite{Miller:2007ri,Loizides:2025ule} to determine the fraction of the total inelastic cross section above a given \etsumhf threshold. 
The fit starts from a minimum \etsumhf corresponding to 80\% centrality, while the contamination due to electromagnetic and diffractive processes from the more peripheral region is studied as a contribution to the systematic uncertainties.

The analysis techniques for two- and multi-particle correlations follow Refs.~\cite{Chatrchyan:2011eka,Chatrchyan:2012wg,CMS:2012qk,CMS:2013jlh,Khachatryan:2014jra,Khachatryan:2015lva,CMS:2015yux}. 
The high-purity track selection criteria described in Ref. \cite{CMS:2014pgm} are used to select primary tracks originating from the primary vertex. 
Additional requirements are also applied to enhance the
purity of primary tracks. The significance of the separation along the beam
axis ($z$) between the track and the primary vertex, $d_z/\sigma(d_z)$, and the significance
of the impact parameter relative to the primary vertex transverse to the beam,
$d_{xy}/\sigma(d_{xy})$, must be smaller than 3.
In the two-particle analysis, for each centrality bin, ``trigger'' particles are defined as
charged particles with $\abs{\eta}<2.4$ and $0.3<\pt<3\GeV$. The number of trigger particles in the event is
denoted by $N_\text{trig}$. Particle pairs are then formed by associating each
trigger particle with the remaining charged particles in the same range of $\abs{\eta}<2.4$ and $0.3<\pt<3\GeV$.
The two-dimensional (2D) per-trigger-particle associated yield is defined as
\begin{equation}
\label{2pcorr_incl}
\frac{1}{N_\text{trig}}\frac{\rd^{2}N^\text{pair}}{\rd\Delta\eta\, \rd\Delta\phi}
= B(0,0)\,\frac{S(\Delta\eta,\Delta\phi)}{B(\Delta\eta,\Delta\phi)},
\end{equation}
where $N^\text{pair}$ is the number of particle pairs and $\Delta\eta$ ($\Delta\phi$) is the difference in $\eta$
($\phi$) of the pair. 
The same-event pair distribution, $S(\Delta\eta,\Delta\phi)$,
represents the yield of particle pairs normalized by $N_\text{trig}$ from the same event.
The mixed-event pair distribution, $B(\Delta\eta,\Delta\phi)$,
is constructed by pairing the trigger particles in each
event with the associated charged particles from 20 different, randomly selected
events in the same 0.5\unit{cm} wide $z_\mathrm{vtx}$ range and from the same centrality class. 
The ratio $B(0,0)/B(\Delta\eta,\Delta\phi)$ mainly accounts for the pair acceptance
effects, with $B(0,0)$ representing the mixed-event associated yield for
both particles of the pair going in approximately the same direction and
thus having maximum pair acceptance.
Each track is weighted by a correction factor derived from the \HIJING v1.383 generator~\cite{Wang:1991hta} 
with \GEANTfour~\cite{Agostinelli:2002hh} used to simulate the CMS detector response, 
following the procedure described in Refs.~\cite{Chatrchyan:2011eka,Chatrchyan:2012wg,CMS:2012qk,CMS:2013jlh,Khachatryan:2014jra,Khachatryan:2015lva,CMS:2015yux}. 
The track weight accounts for detector effects that include the
reconstruction efficiency, the detector acceptance, the fraction of misreconstructed tracks, and the secondary track contamination.

The azimuthal anisotropy harmonics of charged particles can be
extracted via a Fourier decomposition of long-range two-particle \dphi correlation functions. 
These are obtained by averaging the 2D two-particle correlation function over $\abs{\deta}>2$
to remove short-range correlations, such as those from jet fragmentation or resonance decays, using 
\begin{linenomath}
\begin{equation}
\label{eq:Vn}
\frac{1}{N_\text{trig}}\frac{\rd N^\text{pair}}{\rd\Delta\phi} = \frac{N_\text{assoc}}{2\pi} \left[ 1+\sum\limits_{n} 2V_{n\Delta} \cos (n\Delta\phi)\right],
\end{equation}
\end{linenomath}
where $V_{n\Delta}$ are the Fourier coefficients and $N_\text{assoc}$
represents the average number of pairs per trigger particle. 
The first four Fourier terms are included
in the fits to the two-particle correlation functions. Including additional terms has a
negligible effect on the results of the Fourier fit. 
Assuming that the $V_{n\Delta}$ coefficients can be factorized into the product of the single-particle anisotropies \cite{CMS:2013jlh},
the elliptic and triangular anisotropy harmonics, \vsectwo and \vtrdtwo, of trigger particles can be extracted as $\vnn=\sqrt{\smash[b]{V_{n\Delta}}}$. 

The residual contributions of back-to-back dijet correlations 
to the measured \vnn results are estimated using correlations from peripheral events, 
following exactly the procedure established in Refs.~\cite{CMS:2013jlh,Khachatryan:2016txc}. 
The Fourier coefficients, $V_{n\Delta}$, extracted from events with 70--80\% centrality are subtracted from those extracted from
events with 0--50\% centrality after accounting for the jet yield ratio of the selected events, assuming the jet shape does not change from peripheral to central collisions.

The multi-particle analysis is based on correlations among four particles and uses
the $Q$-cumulant method~\cite{Bilandzic:2010jr},
as previously done in proton-lead and lead-lead (\PbPb) collisions~\cite{CMS:2013jlh,CMS:2015yux}.
By simultaneously correlating four
particles, the technique has the advantage of suppressing
short-range, two-particle correlations arising from jets and resonance decays. 
The four-particle correlations also differ in their event-by-event geometry fluctuations response as compared to two-particle measurements. 
In the $Q$-cumulant analysis, two- and four-particle azimuthal correlations~\cite{Bilandzic:2010jr} are evaluated as
\begin{equation}\begin{split}
\dmean{2} &\equiv
	\bigl<\!\bigl< \re^{in(\phi_{1} - \phi_{2})} \bigr>\!\bigr>\quad\text{and}
\\
\dmean{4} &\equiv
    \bigl<\!\bigl< \re^{in(\phi_{1} + \phi_{2} - \phi_{3} - \phi_{4})} \bigr>\!\bigr>,
\label{eq:corr}
\end{split}\end{equation}
where $\phi_{i}$ $(i=1,\ldots,4)$ are the
azimuthal angles of one unique combination of multiple particles in an event, $n$ is the harmonic number,
and $\bigl\langle\bigl\langle \cdots \bigr\rangle\bigr\rangle$ represents
the average over all combinations from all events within a given multiplicity range.
The corresponding cumulants, $\cn{4}$, are calculated as $\cn{4} = \dmean{4} - 2 \dmean{2}^2$. 
The Fourier harmonics $v_n\{4\}$ are related to the cumulants using $\vn{4} = \sqrt[4]{\smash[b]{-\cn{4}}}$.

The systematic uncertainties in the \vnn measurements are estimated by 
exploring the sensitivity of the results to the primary vertex position, 
by varying the track quality requirements,
by estimating the effects from the centrality definition uncertainties, 
and by validating the \vnn extraction methods in simulated samples. 
To evaluate the uncertainties from the vertex dependent detector acceptance, the \vnn values are extracted from events with $\abs{\zvtx}<3\cm$ and with $3<\abs{\zvtx}<15\cm$. 
Experimental systematic uncertainties due to track quality requirements are
examined by varying the track selection thresholds for $d_z/\sigma(d_z)$ and $d_{xy}/\sigma(d_{xy})$
from 2 to 5.
The systematic uncertainties in the centrality definition are studied by repeating the analysis with different higher and lower starting points for the Glauber fit to the HF \etsumhf distribution.
The systematic uncertainties from the \vnn extraction method are evaluated using the \HIJING v1.383~\cite{Wang:1991hta} and AMPT~\cite{Lin:2004en} event generators 
to compare the \vnn values of generator-level particles with those obtained after simulating the detector response using \GEANTfour~\cite{Agostinelli:2002hh} and then applying the same analysis as done for the experimental data.
Potential pileup effects are studied by comparing the reported results to those obtained using events where more than one reconstructed primary vertex is allowed. 
During the data taking, the O and Ne nuclei in the ion beams can fragment into lighter nuclei accompanied by electromagnetic radiation~\cite{Pshenichnov:1997un}. 
This transmutation accumulates as a function of the beam operating time. 
The effects from beam transmutation are found to be negligible by comparing results at the beginning and the end of collider beam fills~\cite{Das:2025rqb}. 
The various sources of systematic uncertainties are added in quadrature to arrive at the final systematic uncertainties.
For the \vsectwo results in \OO\ and \NeNe\ collisions, the systematic uncertainties are found to range from 1.1 to 1.5\% and from 2.1 to 2.3\%, respectively. 
The corresponding uncertainties for the \vtrdtwo results range from 1.8 to 4.5\% and from 1.7 to 4.2\% in \OO and \NeNe collisions, respectively. 
For the \vsecfour results in \OO collisions, a systematic uncertainty of 2.5\% is obtained, while the corresponding uncertainties in \NeNe collisions range from 2.1 to 2.3\%. 
For the ratios of  \NeNe to \OO \vnn values, systematic uncertainties that are correlated between the two systems largely cancel. 
The systematic uncertainties for the ratios then range from 0.9 to 1.8\% for \vsectwo, 
from 1.1 to 1.7\% for \vtrdtwo, and from 0.9 to 1.1\% for \vsecfour.
The total systematic uncertainties of the measurements are found to decrease from peripheral to central collisions. 

\begin{figure}[htb]
\centering
\includegraphics[width=\linewidth]{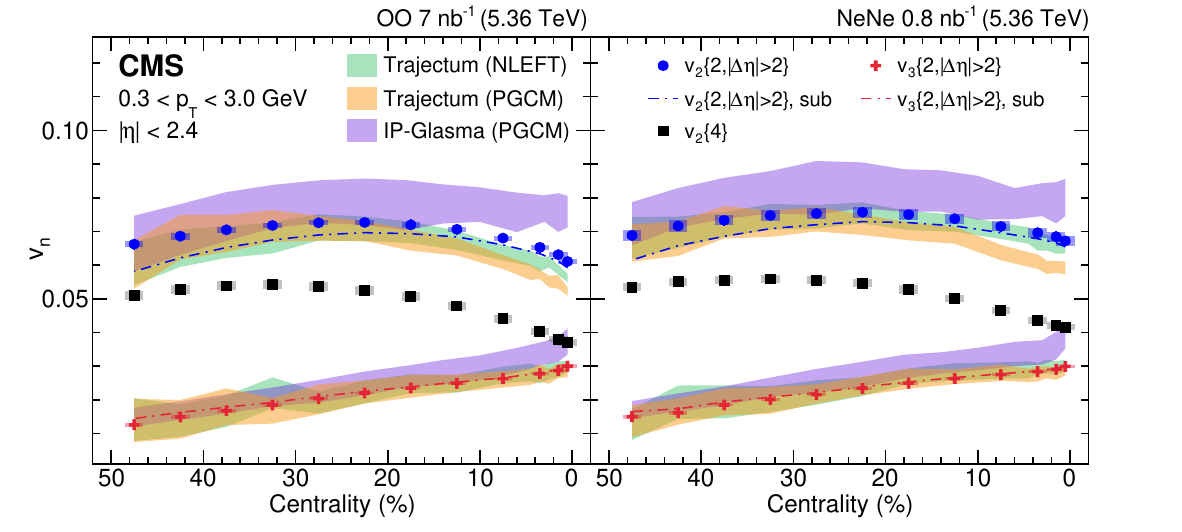}
\caption{The \vsectwo, \vtrdtwo and \vsecfour values for charged particles with $0.3<\pt<3.0\GeV$ and $\abs{\eta}<2.4$ 
are shown as functions of centrality in \OO (left) and \NeNe (right) collisions at $\sqrtsNN=5.36\TeV$. 
The \vsectwo and \vtrdtwo results after subtracting the dijet correlations (denoted as "sub") are indicated by the dot-dashed lines.
Statistical uncertainties are smaller than the markers. The shaded boxes denote systematic uncertainties.
Model calculations~\cite{Giacalone:2024luz,Mantysaari:2025tcg} are shown for comparison to the two-particle correlation results, where the bands represent the combined statistical and systematic uncertainties. 
}
\label{fig:Fig1_vn_centrality}
\end{figure}

The \vsectwo, \vtrdtwo, and \vsecfour results as functions of centrality 
in \OO and \NeNe collisions at $\sqrtsNN=5.36\TeV$ are shown in Fig.~\ref{fig:Fig1_vn_centrality}. 
Dot-dashed lines indicate the results after correcting for dijet correlations. 
This correction decreases from peripheral to central collisions, reducing the \vsectwo and \vtrdtwo values
by 4 to 15\%. 

The \vsectwo values increase from peripheral to mid-central collisions and then decrease in the most central events, 
similarly to those in \PbPb collisions~\cite{CMS:2013wjq,CMS:2019cyz}, reflecting the initial geometry of the overlapping region and the hydrodynamic response of the system. 
The \vtrdtwo values increase toward central collisions. 
While this trend differs from that observed in \PbPb collisions over a wider centrality range, 
it is consistent with the peripheral \PbPb results where the system size is comparable~\cite{CMS:2019wiy}. 
The rise of \vtrd toward more central \OO and \NeNe collisions is therefore attributed to having a stronger hydrodynamic response as the overlap volume increases, 
rather than reflecting a fundamentally different pattern of initial-state fluctuations~\cite{Loizides:2025ule}.
The \vsecfour values are systematically smaller than the \vsectwo values across the shown centrality range.  
This is expected because of different contributions to \vsectwo and \vsecfour from the event-by-event geometry fluctuations~\cite{Ollitrault:2009ie}. 

Model calculations are also shown in Fig.~\ref{fig:Fig1_vn_centrality}. 
The ``\textsc{Trajectum}'' hydrodynamic framework~\cite{Nijs:2020roc,Nijs:2021clz} uses two \textit{ab initio} nuclear structure models, calculated with nuclear lattice effective field theory (NLEFT)~\cite{Lee:2008fa,Lahde:2019npb,Lee:2020meg} 
and with the projected generator coordinate method (PGCM)~\cite{Frosini:2021tuj,Yao:2019rck}, to determine the spatial configurations of nucleons within the O and Ne nuclei. 
The initial energy densities are generated with Trento~\cite{Moreland:2014oya} and evolved with the \textsc{Trajectum} hydrodynamic framework. 
The NLEFT-based calculations reproduce \vsectwo in central collisions better than PGCM, while both agree with data in more peripheral events. 
Both models describe the \vtrdtwo results across the measured centrality range. 
The ``IP-Glasma'' model integrates PGCM into IP-Glasma with JIMWLK evolution~\cite{Mantysaari:2025tcg}, 
followed by MUSIC hydrodynamics~\cite{Schenke:2010nt} and UrQMD hadronic interactions~\cite{Bass:1998ca}. 
It tends to overpredict the \vsectwo and \vtrdtwo values in central collisions, but agrees with data within uncertainties in more peripheral events. 
These comparisons demonstrate that precise \vnn measurements strongly constrain both the initial-state geometry and hydrodynamic response in small collision systems.

\begin{figure}[htb]
\centering
\includegraphics[width=\linewidth]{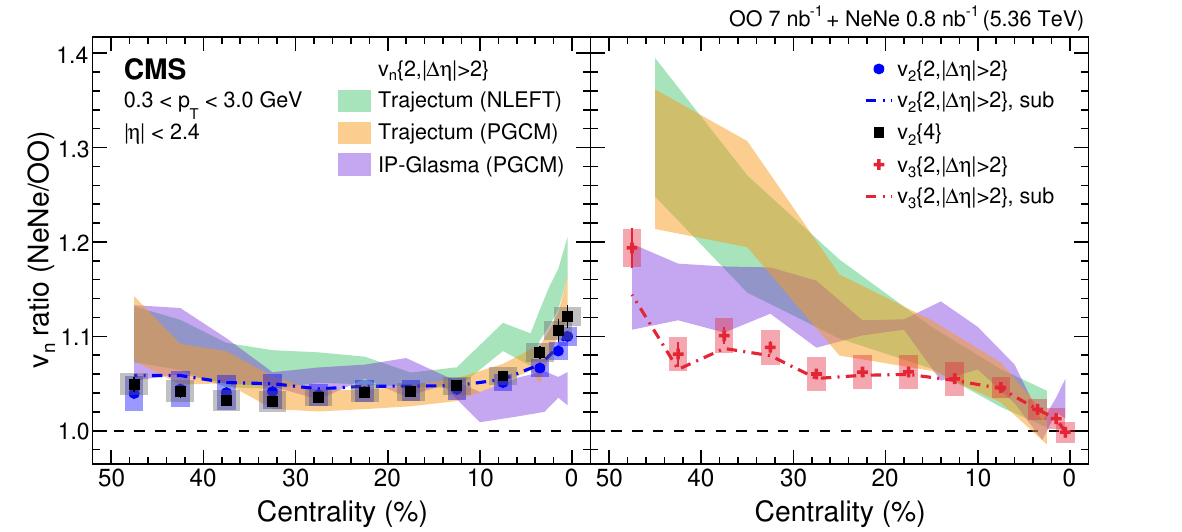}
\caption{Ratios of \vsectwo, \vsecfour (left), and \vtrdtwo (right) of \NeNe to \OO collisions as functions of centrality. 
	Dot-dashed lines indicate results after subtracting (denoted as "sub") the dijet correlations. 
	Vertical bars denote statistical uncertainties, and shaded boxes denote systematic uncertainties.
Model calculations~\cite{Giacalone:2024luz,Mantysaari:2025tcg} are shown for comparison to the two-particle correlation results, where the bands represent the combined statistical and systematic uncertainties. }
\label{fig:Fig2_vn_centrality_ratio}
\end{figure}

To isolate initial-geometry effects, the ratios of  \NeNe to \OO \vnn values are shown in Fig.~\ref{fig:Fig2_vn_centrality_ratio}. 
The results remain consistent, within uncertainties, before and after correcting for dijet correlations. 
A significant rise of the \vsec ratios from two- and four-particle correlations is observed for the most central collisions, while the \vtrd ratios from two-particle correlations exhibit a decreasing trend, which is most pronounced in central events. 
The $^{20}$Ne nucleus is predicted to exhibit stronger intrinsic spatial correlations, resulting in a less spherical initial matter distribution in the central collisions compared to the $^{16}$O nucleus~\cite{Robson:1979zz,PhysRevLett.49.1119,Bijker:2014tka,Marcos:1983msl,Robledo:2011nf,Marevic:2018crl,Lim:2018huo}. 
Therefore, the observed trend of the \OO to \NeNe \vnn ratios provides a sensitive probe of how collective phenomena respond to nuclear structure inputs.
Unlike the conclusions reached when examining individual \vnn values, the Trajectum PGCM calculations reproduce the \vsec ratios more closely, while the NLEFT version tends to overestimate them. 
Both frameworks qualitatively capture the observed increase towards central events, consistent with the expectations based on a deformed geometry of $^{20}$Ne~\cite{Giacalone:2024luz}. 
A quantitative determination of the deformation, however, requires further dedicated studies, since the agreement with data is not uniform across the measured centrality range. 
None of the three model calculations reproduce the measured \vtrd ratios quantitatively, highlighting the need for an improved treatment of initial-state fluctuations in small collision systems. 
The IP-Glasma calculations predict a weaker centrality dependence of the \vsec ratio, likely a result of large multiplicity fluctuations 
that reduce the correlation between centrality and impact parameter~\cite{Mantysaari:2025tcg}. 
Overall, the precise \vnn ratios provide stringent constraints on the modeling of nuclear structure and its implementation in initial-state frameworks, while underscoring the sensitivity of collective flow to the underlying geometry.

In summary, the azimuthal anisotropy of charged-particle emission is measured in oxygen-oxygen (\OO) and neon-neon (\NeNe) collisions at a center-of-mass energy per nucleon pair of 5.36\TeV with the CMS detector.  
The elliptic (\vsec) and triangular (\vtrd) flow harmonics are extracted from two-particle correlations with a large pseudorapidity gap for multiple centrality ranges, 
as defined by the fraction of the total inelastic cross section, from 50\% (peripheral) to 0\% (central) collisions.
The \vsec values are also found using four-particle correlations to highlight how event-by-event geometry fluctuations influence the results.   
The \vsec values rise from peripheral to mid-central events and decrease toward the most central collisions, and the \vtrd values increase monotonically, reflecting hydrodynamic response to the initial geometry.
The ratios of the  \NeNe and \OO \vnn values further isolate initial-geometry effects, with the  \vsec ratio increasing strongly toward central collisions, 
qualitatively consistent with the $^{20}$Ne nucleus being deformed.   
The  \vtrd ratio is found to decrease from peripheral to the most central collisions.
Hydrodynamic models with $ab~initio$ nuclear structure inputs capture the qualitative trends of both the \vnn values and their respective \NeNe to \OO ratios, 
although they fail to fully describe the \vnn magnitude. 
These results establish the sensitivity of collective flow in light ion collisions to both the initial geometry and the hydrodynamic medium response. 
They provide stringent constraints on models of small-system collectivity and offer valuable implications for nuclear structure studies.

\begin{acknowledgments}
We congratulate our colleagues in the CERN accelerator departments for the excellent performance of the LHC and thank the technical and administrative staffs at CERN and at other CMS institutes for their contributions to the success of the CMS effort. In addition, we gratefully acknowledge the computing centers and personnel of the Worldwide LHC Computing Grid and other centers for delivering so effectively the computing infrastructure essential to our analyses. Finally, we acknowledge the enduring support for the construction and operation of the LHC, the CMS detector, and the supporting computing infrastructure provided by the following funding agencies: SC (Armenia), BMBWF and FWF (Austria); FNRS and FWO (Belgium); CNPq, CAPES, FAPERJ, FAPERGS, and FAPESP (Brazil); MES and BNSF (Bulgaria); CERN; CAS, MoST, and NSFC (China); MINCIENCIAS (Colombia); MSES and CSF (Croatia); RIF (Cyprus); SENESCYT (Ecuador); ERC PRG, TARISTU24-TK10 and MoER TK202 (Estonia); Academy of Finland, MEC, and HIP (Finland); CEA and CNRS/IN2P3 (France); SRNSF (Georgia); BMFTR, DFG, and HGF (Germany); GSRI (Greece); NKFIH (Hungary); DAE and DST (India); IPM (Iran); SFI (Ireland); INFN (Italy); MSIT and NRF (Republic of Korea); MES (Latvia); LMTLT (Lithuania); MOE and UM (Malaysia); BUAP, CINVESTAV, CONACYT, LNS, SEP, and UASLP-FAI (Mexico); MOS (Montenegro); MBIE (New Zealand); PAEC (Pakistan); MES, NSC, and NAWA (Poland); FCT (Portugal); MESTD (Serbia); MICIU/AEI and PCTI (Spain); MOSTR (Sri Lanka); Swiss Funding Agencies (Switzerland); MST (Taipei); MHESI (Thailand); TUBITAK and TENMAK (T\"{u}rkiye); NASU (Ukraine); STFC (United Kingdom); DOE and NSF (USA).
\end{acknowledgments}

\bibliography{auto_generated}

\providecommand{\href}[2]{#2}\begingroup\raggedright\begin{thebibliography}{10}%
\makeatletter
\providecommand{\hrefCMSnoop }[0]{\@secondoftwo}%
\makeatother
\providecommand{\doi}{\texttt{doi:}\begingroup \urlstyle{tt}\Url}

\bibitem{Shuryak:1977ut}
\hrefCMSnoop {}{E.~V. Shuryak, ``Theory of hadronic plasma'',} \textit{ Sov.
  Phys. JETP} \textbf{ 47} (1978) 212.

\bibitem{Collins:1974ky}
\hrefCMSnoop {}{J.~C. Collins and M.~J. Perry, ``Superdense matter: Neutrons or
  asymptotically free quarks?'',} \textit{ Phys. Rev. Lett.} \textbf{ 34}
  (1975) 1353,
  \href{http://dx.doi.org/10.1103/PhysRevLett.34.1353}{\doi{10.1103/PhysRevLett.34.1353}}.

\bibitem{STAR:2005ryu}
\hrefCMSnoop {}{{STAR} Collaboration, ``Distributions of charged hadrons
  associated with high transverse momentum particles in pp and {Au+Au}
  collisions at ${\sqrtsNN}$ = 200{\GeV}'',} \textit{ Phys. Rev. Lett.}
  \textbf{ 95} (2005) 152301,
  \href{http://dx.doi.org/10.1103/PhysRevLett.95.152301}{\doi{10.1103/PhysRevLett.95.152301}},
  \href{http://www.arXiv.org/abs/nucl-ex/0501016}{\texttt{arXiv:nucl-ex/0501016}}.

\bibitem{STAR:2009ngv}
\hrefCMSnoop {}{{STAR} Collaboration, ``Long range rapidity correlations and
  jet production in high energy nuclear collisions'',} \textit{ Phys. Rev. C}
  \textbf{ 80} (2009) 064912,
  \href{http://dx.doi.org/10.1103/PhysRevC.80.064912}{\doi{10.1103/PhysRevC.80.064912}},
  \href{http://www.arXiv.org/abs/0909.0191}{\texttt{arXiv:0909.0191}}.

\bibitem{PHOBOS:2009sau}
\hrefCMSnoop {}{{PHOBOS} Collaboration, ``High transverse momentum triggered
  correlations over a large pseudorapidity acceptance in {Au+Au} collisions at
  ${\sqrtsNN}$ = 200{\GeV}'',} \textit{ Phys. Rev. Lett.} \textbf{ 104} (2010)
  062301,
  \href{http://dx.doi.org/10.1103/PhysRevLett.104.062301}{\doi{10.1103/PhysRevLett.104.062301}},
  \href{http://www.arXiv.org/abs/0903.2811}{\texttt{arXiv:0903.2811}}.

\bibitem{Chatrchyan:2011eka}
\hrefCMSnoop {}{{CMS Collaboration}, ``Long-range and short-range dihadron
  angular correlations in central {PbPb} collisions at a nucleon-nucleon center
  of mass energy of 2.76{\TeV}'',} \textit{ JHEP} \textbf{ 07} (2011) 076,
  \href{http://dx.doi.org/10.1007/JHEP07(2011)076}{\doi{10.1007/JHEP07(2011)076}},
\href{http://www.arXiv.org/abs/1105.2438}{\texttt{arXiv:1105.2438}}.

\bibitem{Chatrchyan:2012wg}
\hrefCMSnoop {}{{CMS Collaboration}, ``Centrality dependence of dihadron
  correlations and azimuthal anisotropy harmonics in {PbPb} collisions at
  ${\sqrtsNN}=2.76${\TeV}'',} \textit{ Eur. Phys. J. C} \textbf{ 72} (2012)
  2012,
  \href{http://dx.doi.org/10.1140/epjc/s10052-012-2012-3}{\doi{10.1140/epjc/s10052-012-2012-3}},
\href{http://www.arXiv.org/abs/1201.3158}{\texttt{arXiv:1201.3158}}.

\bibitem{ALICE:2010suc}
\hrefCMSnoop {}{{ALICE Collaboration}, ``Elliptic flow of charged particles in
  {Pb-Pb} collisions at 2.76{\TeV}'',} \textit{ Phys. Rev. Lett.} \textbf{ 105}
  (2010) 252302,
  \href{http://dx.doi.org/10.1103/PhysRevLett.105.252302}{\doi{10.1103/PhysRevLett.105.252302}},
  \href{http://www.arXiv.org/abs/1011.3914}{\texttt{arXiv:1011.3914}}.

\bibitem{ATLAS:2012at}
\hrefCMSnoop {}{{ATLAS Collaboration}, ``Measurement of the azimuthal
  anisotropy for charged particle production in ${\sqrtsNN}=2.76${\TeV}
  lead-lead collisions with the {ATLAS} detector'',} \textit{ Phys. Rev. C}
  \textbf{ 86} (2012) 014907,
  \href{http://dx.doi.org/10.1103/PhysRevC.86.014907}{\doi{10.1103/PhysRevC.86.014907}},
  \href{http://www.arXiv.org/abs/1203.3087}{\texttt{arXiv:1203.3087}}.

\bibitem{CMS:2012zex}
\hrefCMSnoop {}{{CMS Collaboration}, ``Measurement of the elliptic anisotropy
  of charged particles produced in {PbPb} collisions at
  ${\sqrtsNN}=2.76${\TeV}'',} \textit{ Phys. Rev. C} \textbf{ 87} (2013)
  014902,
  \href{http://dx.doi.org/10.1103/PhysRevC.87.014902}{\doi{10.1103/PhysRevC.87.014902}},
  \href{http://www.arXiv.org/abs/1204.1409}{\texttt{arXiv:1204.1409}}.

\bibitem{CMS:2013bza}
\hrefCMSnoop {}{{CMS Collaboration}, ``Studies of azimuthal dihadron
  correlations in ultra-central {PbPb} collisions at ${\sqrtsNN} =$
  2.76{\TeV}'',} \textit{ JHEP} \textbf{ 02} (2014) 088,
  \href{http://dx.doi.org/10.1007/JHEP02(2014)088}{\doi{10.1007/JHEP02(2014)088}},
  \href{http://www.arXiv.org/abs/1312.1845}{\texttt{arXiv:1312.1845}}.

\bibitem{CMS:2024krd}
\hrefCMSnoop {}{{CMS Collaboration}, ``{Overview of high-density QCD studies
  with the CMS experiment at the LHC}'',} \textit{ Phys. Rept.} \textbf{ 1115}
  (2025) 219,
  \href{http://dx.doi.org/10.1016/j.physrep.2024.11.007}{\doi{10.1016/j.physrep.2024.11.007}},
  \href{http://www.arXiv.org/abs/2405.10785}{\texttt{arXiv:2405.10785}}.

\bibitem{Ollitrault:1992bk}
\hrefCMSnoop {}{J.-Y. Ollitrault, ``Anisotropy as a signature of transverse
  collective flow'',} \textit{ Phys. Rev. D} \textbf{ 46} (1992) 229,
\href{http://dx.doi.org/10.1103/PhysRevD.46.229}{\doi{10.1103/PhysRevD.46.229}}.

\bibitem{Heinz:2013th}
\hrefCMSnoop {}{U.~Heinz and R.~Snellings, ``Collective flow and viscosity in
  relativistic heavy-ion collisions'',} \textit{ Ann. Rev. Nucl. Part. Sci.}
  \textbf{ 63} (2013) 123,
  \href{http://dx.doi.org/10.1146/annurev-nucl-102212-170540}{\doi{10.1146/annurev-nucl-102212-170540}},
\href{http://www.arXiv.org/abs/1301.2826}{\texttt{arXiv:1301.2826}}.

\bibitem{Gale:2013da}
\hrefCMSnoop {}{C.~Gale, S.~Jeon, and B.~Schenke, ``Hydrodynamic modeling of
  heavy ion collisions'',} \textit{ Int. J. Mod. Phys. A} \textbf{ 28} (2013)
  1340011,
  \href{http://dx.doi.org/10.1142/S0217751X13400113}{\doi{10.1142/S0217751X13400113}},
\href{http://www.arXiv.org/abs/1301.5893}{\texttt{arXiv:1301.5893}}.

\bibitem{Alver:2010dn}
\hrefCMSnoop {}{B.~H. Alver, C.~Gombeaud, M.~Luzum, and J.-Y. Ollitrault,
  ``Triangular flow in hydrodynamics and transport theory'',} \textit{ Phys.
  Rev. C} \textbf{ 82} (2010) 034913,
  \href{http://dx.doi.org/10.1103/PhysRevC.82.034913}{\doi{10.1103/PhysRevC.82.034913}},
  \href{http://www.arXiv.org/abs/1007.5469}{\texttt{arXiv:1007.5469}}.

\bibitem{Schenke:2010rr}
\hrefCMSnoop {}{B.~Schenke, S.~Jeon, and C.~Gale, ``Elliptic and triangular
  flow in event-by-event {(3+1)D} viscous hydrodynamics'',} \textit{ Phys. Rev.
  Lett.} \textbf{ 106} (2011) 042301,
  \href{http://dx.doi.org/10.1103/PhysRevLett.106.042301}{\doi{10.1103/PhysRevLett.106.042301}},
  \href{http://www.arXiv.org/abs/1009.3244}{\texttt{arXiv:1009.3244}}.

\bibitem{Qiu:2011hf}
\hrefCMSnoop {}{Z.~Qiu, C.~Shen, and U.~Heinz, ``Hydrodynamic elliptic and
  triangular flow in {Pb-Pb} collisions at {$\sqrt{s}=2.76$}{ATeV}'',} \textit{
  Phys. Lett. B} \textbf{ 707} (2012) 151,
  \href{http://dx.doi.org/10.1016/j.physletb.2011.12.041}{\doi{10.1016/j.physletb.2011.12.041}},
  \href{http://www.arXiv.org/abs/1110.3033}{\texttt{arXiv:1110.3033}}.

\bibitem{Khachatryan:2010gv}
\hrefCMSnoop {}{{CMS Collaboration}, ``Observation of long-range near-side
  angular correlations in proton-proton collisions at the {LHC}'',} \textit{
  JHEP} \textbf{ 09} (2010) 091,
  \href{http://dx.doi.org/10.1007/JHEP09(2010)091}{\doi{10.1007/JHEP09(2010)091}},
\href{http://www.arXiv.org/abs/1009.4122}{\texttt{arXiv:1009.4122}}.

\bibitem{Aad:2015gqa}
\hrefCMSnoop {}{{ATLAS Collaboration}, ``Observation of long-range elliptic
  azimuthal anisotropies in $\sqrt{s}=13$ and 2.76{\TeV} \pp collisions with
  the {ATLAS} detector'',} \textit{ Phys. Rev. Lett.} \textbf{ 116} (2016)
  172301,
  \href{http://dx.doi.org/10.1103/PhysRevLett.116.172301}{\doi{10.1103/PhysRevLett.116.172301}},
  \href{http://www.arXiv.org/abs/1509.04776}{\texttt{arXiv:1509.04776}}.

\bibitem{Khachatryan:2015lva}
\hrefCMSnoop {}{{CMS Collaboration}, ``Measurement of long-range near-side
  two-particle angular correlations in \pp collisions at $\sqrt{s}=
  13${\TeV}'',} \textit{ Phys. Rev. Lett.} \textbf{ 116} (2016) 172302,
  \href{http://dx.doi.org/10.1103/PhysRevLett.116.172302}{\doi{10.1103/PhysRevLett.116.172302}},
  \href{http://www.arXiv.org/abs/1510.03068}{\texttt{arXiv:1510.03068}}.

\bibitem{Khachatryan:2016txc}
\hrefCMSnoop {}{{CMS Collaboration}, ``Evidence for collectivity in \pp
  collisions at the {LHC}'',} \textit{ Phys. Lett. B} \textbf{ 765} (2017) 193,
  \href{http://dx.doi.org/10.1016/j.physletb.2016.12.009}{\doi{10.1016/j.physletb.2016.12.009}},
\href{http://www.arXiv.org/abs/1606.06198}{\texttt{arXiv:1606.06198}}.

\bibitem{CMS:2020qul}
\hrefCMSnoop {}{{CMS Collaboration}, ``Studies of charm and beauty hadron
  long-range correlations in \pp and {\pPb} collisions at {LHC} energies'',}
  \textit{ Phys. Lett. B} \textbf{ 813} (2021) 136036,
  \href{http://dx.doi.org/10.1016/j.physletb.2020.136036}{\doi{10.1016/j.physletb.2020.136036}},
  \href{http://www.arXiv.org/abs/2009.07065}{\texttt{arXiv:2009.07065}}.

\bibitem{CMS:2012qk}
\hrefCMSnoop {}{{CMS Collaboration}, ``Observation of long-range near-side
  angular correlations in proton-lead collisions at the {LHC}'',} \textit{
  Phys. Lett. B} \textbf{ 718} (2013) 795,
  \href{http://dx.doi.org/10.1016/j.physletb.2012.11.025}{\doi{10.1016/j.physletb.2012.11.025}},
\href{http://www.arXiv.org/abs/1210.5482}{\texttt{arXiv:1210.5482}}.

\bibitem{alice:2012qe}
\hrefCMSnoop {}{{ALICE Collaboration}, ``Long-range angular correlations on the
  near and away side in {\pPb} collisions at ${\sqrtsNN}=5.02{\TeV}$'',}
  \textit{ Phys. Lett. B} \textbf{ 719} (2013) 29,
  \href{http://dx.doi.org/10.1016/j.physletb.2013.01.012}{\doi{10.1016/j.physletb.2013.01.012}},
\href{http://www.arXiv.org/abs/1212.2001}{\texttt{arXiv:1212.2001}}.

\bibitem{Aad:2012gla}
\hrefCMSnoop {}{{ATLAS Collaboration}, ``Observation of associated near-side
  and away-side long-range correlations in ${\sqrtsNN}=5.02{\TeV}$ proton-lead
  collisions with the {ATLAS} detector'',} \textit{ Phys. Rev. Lett.} \textbf{
  110} (2013) 182302,
  \href{http://dx.doi.org/10.1103/PhysRevLett.110.182302}{\doi{10.1103/PhysRevLett.110.182302}},
\href{http://www.arXiv.org/abs/1212.5198}{\texttt{arXiv:1212.5198}}.

\bibitem{LHCb:2015coe}
\hrefCMSnoop {}{{LHCb Collaboration}, ``Measurements of long-range near-side
  angular correlations in ${\sqrtsNN}=5{\TeV}$ proton-lead collisions in the
  forward region'',} \textit{ Phys. Lett. B} \textbf{ 762} (2016) 473,
  \href{http://dx.doi.org/10.1016/j.physletb.2016.09.064}{\doi{10.1016/j.physletb.2016.09.064}},
  \href{http://www.arXiv.org/abs/1512.00439}{\texttt{arXiv:1512.00439}}.

\bibitem{CMS:2015yux}
\hrefCMSnoop {}{{CMS Collaboration}, ``Evidence for collective multiparticle
  correlations in {\pPb} collisions'',} \textit{ Phys. Rev. Lett.} \textbf{
  115} (2015) 012301,
  \href{http://dx.doi.org/10.1103/PhysRevLett.115.012301}{\doi{10.1103/PhysRevLett.115.012301}},
  \href{http://www.arXiv.org/abs/1502.05382}{\texttt{arXiv:1502.05382}}.

\bibitem{CMS:2018loe}
\hrefCMSnoop {}{{CMS Collaboration}, ``Elliptic flow of charm and strange
  hadrons in high-multiplicity {\pPb} collisions at ${\sqrtsNN}=8.16{\TeV}$'',}
  \textit{ Phys. Rev. Lett.} \textbf{ 121} (2018) 082301,
  \href{http://dx.doi.org/10.1103/PhysRevLett.121.082301}{\doi{10.1103/PhysRevLett.121.082301}},
  \href{http://www.arXiv.org/abs/1804.09767}{\texttt{arXiv:1804.09767}}.

\bibitem{CMS:2018duw}
\hrefCMSnoop {}{{CMS Collaboration}, ``Observation of prompt {J}/$\psi$ meson
  elliptic flow in high-multiplicity {\pPb} collisions at
  ${\sqrtsNN}=8.16{\TeV}$'',} \textit{ Phys. Lett. B} \textbf{ 791} (2019) 172,
  \href{http://dx.doi.org/10.1016/j.physletb.2019.02.018}{\doi{10.1016/j.physletb.2019.02.018}},
  \href{http://www.arXiv.org/abs/1810.01473}{\texttt{arXiv:1810.01473}}.

\bibitem{CMS:2019wiy}
\hrefCMSnoop {}{{CMS Collaboration}, ``Multiparticle correlation studies in
  {\pPb} collisions at {$\sqrtsNN=$}8.16{\TeV}'',} \textit{ Phys. Rev. C}
  \textbf{ 101} (2020) 014912,
  \href{http://dx.doi.org/10.1103/PhysRevC.101.014912}{\doi{10.1103/PhysRevC.101.014912}},
  \href{http://www.arXiv.org/abs/1904.11519}{\texttt{arXiv:1904.11519}}.

\bibitem{CMS:2022bmk}
\hrefCMSnoop {}{{CMS Collaboration}, ``Strange hadron collectivity in {\pPb}
  and {\PbPb} collisions'',} \textit{ JHEP} \textbf{ 05} (2023) 007,
  \href{http://dx.doi.org/10.1007/JHEP05(2023)007}{\doi{10.1007/JHEP05(2023)007}},
  \href{http://www.arXiv.org/abs/2205.00080}{\texttt{arXiv:2205.00080}}.

\bibitem{PHENIX:2014fnc}
\hrefCMSnoop {}{{PHENIX} Collaboration, ``Measurement of long-range angular
  correlation and quadrupole anisotropy of pions and (anti)protons in central
  {$d+$Au} collisions at {$\sqrtsNN=200\GeV$}'',} \textit{ Phys. Rev. Lett.}
  \textbf{ 114} (2015) 192301,
  \href{http://dx.doi.org/10.1103/PhysRevLett.114.192301}{\doi{10.1103/PhysRevLett.114.192301}},
  \href{http://www.arXiv.org/abs/1404.7461}{\texttt{arXiv:1404.7461}}.

\bibitem{PHENIX:2015idk}
\hrefCMSnoop {}{{PHENIX} Collaboration, ``Measurements of elliptic and
  triangular flow in high-multiplicity {$^{3}$He$+$Au} collisions at
  {$\sqrtsNN=200\GeV$}'',} \textit{ Phys. Rev. Lett.} \textbf{ 115} (2015)
  142301,
  \href{http://dx.doi.org/10.1103/PhysRevLett.115.142301}{\doi{10.1103/PhysRevLett.115.142301}},
  \href{http://www.arXiv.org/abs/1507.06273}{\texttt{arXiv:1507.06273}}.

\bibitem{PHENIX:2016cfs}
\hrefCMSnoop {}{{PHENIX} Collaboration, ``Measurement of long-range angular
  correlations and azimuthal anisotropies in high-multiplicity
  {$p+\mathrm{Au}$} collisions at {$\sqrtsNN=200\GeV$}'',} \textit{ Phys. Rev.
  C} \textbf{ 95} (2017) 034910,
  \href{http://dx.doi.org/10.1103/PhysRevC.95.034910}{\doi{10.1103/PhysRevC.95.034910}},
  \href{http://www.arXiv.org/abs/1609.02894}{\texttt{arXiv:1609.02894}}.

\bibitem{PHENIX:2017djs}
\hrefCMSnoop {}{{PHENIX} Collaboration, ``Measurements of mass-dependent
  azimuthal anisotropy in central {$p+\mathrm{Au}$}, {$d+\mathrm{Au}$}, and
  {$^{3}\mathrm{He}+\mathrm{Au}$} collisions at {$\sqrtsNN=200\GeV$}'',}
  \textit{ Phys. Rev. C} \textbf{ 97} (2018) 064904,
  \href{http://dx.doi.org/10.1103/PhysRevC.97.064904}{\doi{10.1103/PhysRevC.97.064904}},
  \href{http://www.arXiv.org/abs/1710.09736}{\texttt{arXiv:1710.09736}}.

\bibitem{PHENIX:2018lia}
\hrefCMSnoop {}{{PHENIX} Collaboration, ``Creation of quark\textendash{}gluon
  plasma droplets with three distinct geometries'',} \textit{ Nature Phys.}
  \textbf{ 15} (2019) 214,
  \href{http://dx.doi.org/10.1038/s41567-018-0360-0}{\doi{10.1038/s41567-018-0360-0}},
  \href{http://www.arXiv.org/abs/1805.02973}{\texttt{arXiv:1805.02973}}.

\bibitem{PHENIX:2021ubk}
\hrefCMSnoop {}{{PHENIX} Collaboration, ``Kinematic dependence of azimuthal
  anisotropies in {$p+\mathrm{Au}$}, {$d+\mathrm{Au}$}, and
  {$^{3}\mathrm{He}+\mathrm{Au}$} at {$\sqrtsNN=200\GeV$}'',} \textit{ Phys.
  Rev. C} \textbf{ 105} (2022) 024901,
  \href{http://dx.doi.org/10.1103/PhysRevC.105.024901}{\doi{10.1103/PhysRevC.105.024901}},
  \href{http://www.arXiv.org/abs/2107.06634}{\texttt{arXiv:2107.06634}}.

\bibitem{STAR:2015kak}
\hrefCMSnoop {}{{STAR} Collaboration, ``Long-range pseudorapidity dihadron
  correlations in {$d$+Au} collisions at {$\sqrtsNN=200\GeV$}'',} \textit{
  Phys. Lett. B} \textbf{ 747} (2015) 265,
  \href{http://dx.doi.org/10.1016/j.physletb.2015.05.075}{\doi{10.1016/j.physletb.2015.05.075}},
  \href{http://www.arXiv.org/abs/1502.07652}{\texttt{arXiv:1502.07652}}.

\bibitem{STAR:2019zaf}
\hrefCMSnoop {}{{STAR} Collaboration, ``{Azimuthal harmonics in small and large
  collision systems at RHIC top energies}'',} \textit{ Phys. Rev. Lett.}
  \textbf{ 122} (2019) 172301,
  \href{http://dx.doi.org/10.1103/PhysRevLett.122.172301}{\doi{10.1103/PhysRevLett.122.172301}},
  \href{http://www.arXiv.org/abs/1901.08155}{\texttt{arXiv:1901.08155}}.

\bibitem{STAR:2022pfn}
\hrefCMSnoop {}{{STAR} Collaboration, ``Measurements of the elliptic and
  triangular azimuthal anisotropies in central {$^{3}\mathrm{He}+\mathrm{Au}$},
  {$d+\mathrm{Au}$} and {$p+\mathrm{Au}$} collisions at
  ${\sqrtsNN}=200{\GeV}$'',} \textit{ Phys. Rev. Lett.} \textbf{ 130} (2023)
  242301,
  \href{http://dx.doi.org/10.1103/PhysRevLett.130.242301}{\doi{10.1103/PhysRevLett.130.242301}},
  \href{http://www.arXiv.org/abs/2210.11352}{\texttt{arXiv:2210.11352}}.

\bibitem{STAR:2023wmd}
\hrefCMSnoop {}{{STAR} Collaboration, ``Measurement of flow coefficients in
  high-multiplicity {$p+\mathrm{Au}$}, {$d+\mathrm{Au}$}, and
  {$^{3}\mathrm{He}+\mathrm{Au}$} collisions at {$\sqrtsNN=200\GeV$}'',}
  \textit{ Phys. Rev. C} \textbf{ 110} (2024) 064902,
  \href{http://dx.doi.org/10.1103/PhysRevC.110.064902}{\doi{10.1103/PhysRevC.110.064902}},
  \href{http://www.arXiv.org/abs/2312.07464}{\texttt{arXiv:2312.07464}}.

\bibitem{Nagle:2018nvi}
\hrefCMSnoop {}{J.~L. Nagle and W.~A. Zajc, ``Small system collectivity in
  relativistic hadronic and nuclear collisions'',} \textit{ Ann. Rev. Nucl.
  Part. Sci.} \textbf{ 68} (2018) 211,
  \href{http://dx.doi.org/10.1146/annurev-nucl-101916-123209}{\doi{10.1146/annurev-nucl-101916-123209}},
  \href{http://www.arXiv.org/abs/1801.03477}{\texttt{arXiv:1801.03477}}.

\bibitem{Dusling:2015gta}
\hrefCMSnoop {}{K.~Dusling, W.~Li, and B.~Schenke, ``Novel collective phenomena
  in high-energy proton-proton and proton-nucleus collisions'',} \textit{ Int.
  J. Mod. Phys. E} \textbf{ 25} (2016) 1630002,
  \href{http://dx.doi.org/10.1142/S0218301316300022}{\doi{10.1142/S0218301316300022}},
\href{http://www.arXiv.org/abs/1509.07939}{\texttt{arXiv:1509.07939}}.

\bibitem{Schenke:2021mxx}
\hrefCMSnoop {}{B.~Schenke, ``The smallest fluid on earth'',} \textit{ Rept.
  Prog. Phys.} \textbf{ 84} (2021) 082301,
  \href{http://dx.doi.org/10.1088/1361-6633/ac14c9}{\doi{10.1088/1361-6633/ac14c9}},
  \href{http://www.arXiv.org/abs/2102.11189}{\texttt{arXiv:2102.11189}}.

\bibitem{Zhao:2022ugy}
\hrefCMSnoop {}{W.~Zhao, S.~Ryu, C.~Shen, and B.~Schenke, ``{3D} structure of
  anisotropic flow in small collision systems at energies available at the {BNL
  Relativistic Heavy Ion Collider}'',} \textit{ Phys. Rev. C} \textbf{ 107}
  (2023) 014904,
  \href{http://dx.doi.org/10.1103/PhysRevC.107.014904}{\doi{10.1103/PhysRevC.107.014904}},
  \href{http://www.arXiv.org/abs/2211.16376}{\texttt{arXiv:2211.16376}}.

\bibitem{Bozek:2015bna}
\hrefCMSnoop {}{P.~Bozek and W.~Broniowski, ``The torque effect and
  fluctuations of entropy deposition in rapidity in ultra-relativistic nuclear
  collisions'',} \textit{ Phys. Lett. B} \textbf{ 752} (2016) 206,
  \href{http://dx.doi.org/10.1016/j.physletb.2015.11.054}{\doi{10.1016/j.physletb.2015.11.054}},
  \href{http://www.arXiv.org/abs/1506.02817}{\texttt{arXiv:1506.02817}}.

\bibitem{Giacalone:2024luz}
\hrefCMSnoop {}{G.~Giacalone { et~al.}, ``Exploiting {$^{20}\mathrm{Ne}$}
  isotopes for precision characterizations of collectivity in small systems'',}
  \textit{ Phys. Rev. Lett.} \textbf{ 135} (2025) 012302,
  \href{http://dx.doi.org/10.1103/k8rb-jgvq}{\doi{10.1103/k8rb-jgvq}},
  \href{http://www.arXiv.org/abs/2402.05995}{\texttt{arXiv:2402.05995}}.

\bibitem{Peter:1980}
P.~Ring and P.~Schuck, ``The Nuclear Many-Body Problem''.
\newblock Springer, 1980.
\newblock ISBN~978-3-540-21206-5.

\bibitem{CMS:2019cyz}
\hrefCMSnoop {}{{CMS Collaboration}, ``Charged-particle angular correlations in
  {XeXe} collisions at ${\sqrtsNN}=$ 5.44{\TeV}'',} \textit{ Phys. Rev. C}
  \textbf{ 100} (2019) 044902,
  \href{http://dx.doi.org/10.1103/PhysRevC.100.044902}{\doi{10.1103/PhysRevC.100.044902}},
  \href{http://www.arXiv.org/abs/1901.07997}{\texttt{arXiv:1901.07997}}.

\bibitem{Li:2019kkh}
H.~Li\hrefCMSnoop {}{ { et~al.}, ``Probing the neutron skin with
  ultrarelativistic isobaric collisions'',} \textit{ Phys. Rev. Lett.} \textbf{
  125} (2020) 222301,
  \href{http://dx.doi.org/10.1103/PhysRevLett.125.222301}{\doi{10.1103/PhysRevLett.125.222301}},
  \href{http://www.arXiv.org/abs/1910.06170}{\texttt{arXiv:1910.06170}}.

\bibitem{Xu:2021uar}
H.-j. Xu\hrefCMSnoop {}{ { et~al.}, ``Probing nuclear structure with mean
  transverse momentum in relativistic isobar collisions'',} \textit{ Phys. Rev.
  C} \textbf{ 108} (2023) L011902,
  \href{http://dx.doi.org/10.1103/PhysRevC.108.L011902}{\doi{10.1103/PhysRevC.108.L011902}},
  \href{http://www.arXiv.org/abs/2111.14812}{\texttt{arXiv:2111.14812}}.

\bibitem{Zhang:2021kxj}
\hrefCMSnoop {}{C.~Zhang and J.~Jia, ``Evidence of quadrupole and octupole
  deformations in {$^{96}\mathrm{Zr}$+$^{96}\mathrm{Zr}$} and
  {$^{96}\mathrm{Ru}$+$^{96}\mathrm{Ru}$} collisions at ultrarelativistic
  energies'',} \textit{ Phys. Rev. Lett.} \textbf{ 128} (2022) 022301,
  \href{http://dx.doi.org/10.1103/PhysRevLett.128.022301}{\doi{10.1103/PhysRevLett.128.022301}},
  \href{http://www.arXiv.org/abs/2109.01631}{\texttt{arXiv:2109.01631}}.

\bibitem{Ryssens:2023fkv}
\hrefCMSnoop {}{W.~Ryssens, G.~Giacalone, B.~Schenke, and C.~Shen, ``Evidence
  of hexadecapole deformation in uranium-238 at the {Relativistic Heavy Ion
  Collider}'',} \textit{ Phys. Rev. Lett.} \textbf{ 130} (2023) 212302,
  \href{http://dx.doi.org/10.1103/PhysRevLett.130.212302}{\doi{10.1103/PhysRevLett.130.212302}},
  \href{http://www.arXiv.org/abs/2302.13617}{\texttt{arXiv:2302.13617}}.

\bibitem{Xu:2024bdh}
\hrefCMSnoop {}{H.-j. Xu, J.~Zhao, and F.~Wang, ``Hexadecapole deformation of
  {$^{238}\mathrm{U}$} from relativistic heavy-ion collisions using a nonlinear
  response coefficient'',} \textit{ Phys. Rev. Lett.} \textbf{ 132} (2024)
  262301,
  \href{http://dx.doi.org/10.1103/PhysRevLett.132.262301}{\doi{10.1103/PhysRevLett.132.262301}},
  \href{http://www.arXiv.org/abs/2402.16550}{\texttt{arXiv:2402.16550}}.

\bibitem{STAR:2024wgy}
\hrefCMSnoop {}{{STAR} Collaboration, ``Imaging shapes of atomic nuclei in
  high-energy nuclear collisions'',} \textit{ Nature} \textbf{ 635} (2024) 67,
  \href{http://dx.doi.org/10.1038/s41586-024-08097-2}{\doi{10.1038/s41586-024-08097-2}},
  \href{http://www.arXiv.org/abs/2401.06625}{\texttt{arXiv:2401.06625}}.

\bibitem{Giacalone:2024ixe}
\hrefCMSnoop {}{G.~Giacalone { et~al.}, ``Anisotropic flow in fixed-target
  {$^{208}\mathrm{Pb}$+$^{20}\mathrm{Ne}$} collisions as a probe of quark-gluon
  plasma'',} \textit{ Phys. Rev. Lett.} \textbf{ 134} (2025) 082301,
  \href{http://dx.doi.org/10.1103/PhysRevLett.134.082301}{\doi{10.1103/PhysRevLett.134.082301}},
  \href{http://www.arXiv.org/abs/2405.20210}{\texttt{arXiv:2405.20210}}.

\bibitem{Robson:1979zz}
\hrefCMSnoop {}{D.~Robson, ``Evidence for the tetrahedral nature of
  {$^{16}\mathrm{O}$}'',} \textit{ Phys. Rev. Lett.} \textbf{ 42} (1979) 876,
  \href{http://dx.doi.org/10.1103/PhysRevLett.42.876}{\doi{10.1103/PhysRevLett.42.876}}.

\bibitem{PhysRevLett.49.1119}
\hrefCMSnoop {}{L.~Pauling, ``Comment on the test for tetrahedral symmetry in
  the {$^{16}\mathrm{O}$} nucleus and its relation to the shell model'',}
  \textit{ Phys. Rev. Lett.} \textbf{ 49} (1982) 1119,
  \href{http://dx.doi.org/10.1103/PhysRevLett.49.1119}{\doi{10.1103/PhysRevLett.49.1119}}.

\bibitem{Bijker:2014tka}
\hrefCMSnoop {}{R.~Bijker and F.~Iachello, ``Evidence for tetrahedral symmetry
  in {$^{16}\mathrm{O}$}'',} \textit{ Phys. Rev. Lett.} \textbf{ 112} (2014)
  152501,
  \href{http://dx.doi.org/10.1103/PhysRevLett.112.152501}{\doi{10.1103/PhysRevLett.112.152501}},
  \href{http://www.arXiv.org/abs/1403.6773}{\texttt{arXiv:1403.6773}}.

\bibitem{Marcos:1983msl}
\hrefCMSnoop {}{S.~Marcos, H.~Flocard, and P.~H. Heenen, ``Influence of
  left-right asymmetry degrees of freedom in self-consistent calculations of
  {$^{20}\mathrm{Ne}$}'',} \textit{ Nucl. Phys. A} \textbf{ 410} (1983) 125,
  \href{http://dx.doi.org/10.1016/0375-9474(83)90405-0}{\doi{10.1016/0375-9474(83)90405-0}}.

\bibitem{Robledo:2011nf}
\hrefCMSnoop {}{L.~M. Robledo and G.~F. Bertsch, ``Global systematics of
  octupole excitations in even-even nuclei'',} \textit{ Phys. Rev. C} \textbf{
  84} (2011) 054302,
  \href{http://dx.doi.org/10.1103/PhysRevC.84.054302}{\doi{10.1103/PhysRevC.84.054302}},
  \href{http://www.arXiv.org/abs/1107.3581}{\texttt{arXiv:1107.3581}}.

\bibitem{Marevic:2018crl}
P.~Marevi{\'c}\hrefCMSnoop {}{ { et~al.}, ``Quadrupole and octupole
  collectivity and cluster structures in neon isotopes'',} \textit{ Phys. Rev.
  C} \textbf{ 97} (2018) 024334,
  \href{http://dx.doi.org/10.1103/PhysRevC.97.024334}{\doi{10.1103/PhysRevC.97.024334}},
  \href{http://www.arXiv.org/abs/1802.02873}{\texttt{arXiv:1802.02873}}.

\bibitem{Lim:2018huo}
S.~H. Lim\hrefCMSnoop {}{ { et~al.}, ``Exploring new small system geometries in
  heavy ion collisions'',} \textit{ Phys. Rev. C} \textbf{ 99} (2019) 044904,
  \href{http://dx.doi.org/10.1103/PhysRevC.99.044904}{\doi{10.1103/PhysRevC.99.044904}},
  \href{http://www.arXiv.org/abs/1812.08096}{\texttt{arXiv:1812.08096}}.

\bibitem{Epelbaum:2013paa}
E.~Epelbaum\hrefCMSnoop {}{ { et~al.}, ``Ab initio calculation of the spectrum
  and structure of {$^{16}\mathrm{O}$}'',} \textit{ Phys. Rev. Lett.} \textbf{
  112} (2014) 102501,
  \href{http://dx.doi.org/10.1103/PhysRevLett.112.102501}{\doi{10.1103/PhysRevLett.112.102501}},
  \href{http://www.arXiv.org/abs/1312.7703}{\texttt{arXiv:1312.7703}}.

\bibitem{Pieper:2002ne}
\hrefCMSnoop {}{S.~C. Pieper, K.~Varga, and R.~B. Wiringa, ``Quantum {Monte
  Carlo} calculations of {A=9, A=10} nuclei'',} \textit{ Phys. Rev. C} \textbf{
  66} (2002) 044310,
  \href{http://dx.doi.org/10.1103/PhysRevC.66.044310}{\doi{10.1103/PhysRevC.66.044310}},
  \href{http://www.arXiv.org/abs/nucl-th/0206061}{\texttt{arXiv:nucl-th/0206061}}.

\bibitem{Lee:2008fa}
\hrefCMSnoop {}{D.~Lee, ``Lattice simulations for few- and many-body
  systems'',} \textit{ Prog. Part. Nucl. Phys.} \textbf{ 63} (2009) 117,
  \href{http://dx.doi.org/10.1016/j.ppnp.2008.12.001}{\doi{10.1016/j.ppnp.2008.12.001}},
  \href{http://www.arXiv.org/abs/0804.3501}{\texttt{arXiv:0804.3501}}.

\bibitem{Dobaczewski:2025rdi}
J.~Dobaczewski\hrefCMSnoop {}{ { et~al.}, ``Extraction of ground-state nuclear
  deformations from ultrarelativistic heavy-ion collisions: Nuclear structure
  physics context'',} \textit{ Phys. Rev. Res.} \textbf{ 7} (2025) 043159,
  \href{http://dx.doi.org/10.1103/kngg-1ccb}{\doi{10.1103/kngg-1ccb}},
  \href{http://www.arXiv.org/abs/2507.05208}{\texttt{arXiv:2507.05208}}.

\bibitem{Hagino:2025vxe}
\hrefCMSnoop {}{K.~Hagino and M.~Kitazawa, ``Probing surface vibrations of
  spherical nuclei in relativistic heavy-ion collisions'',} \textit{ Phys. Rev.
  C} \textbf{ 112} (2025) L041901,
  \href{http://dx.doi.org/10.1103/2f3s-2nyn}{\doi{10.1103/2f3s-2nyn}},
  \href{http://www.arXiv.org/abs/2508.05125}{\texttt{arXiv:2508.05125}}.

\bibitem{hepdata}
\hrefCMSnoop {}{{CMS Collaboration}, ``{HEPD}ata record for this analysis'',}
  2025.
\newblock
  \href{http://dx.doi.org/10.17182/hepdata.165513}{\doi{10.17182/hepdata.165513}}.

\bibitem{ALICE:2025luc}
\hrefCMSnoop {}{{ALICE Collaboration}, ``Evidence of nuclear geometry-driven
  anisotropic flow in {OO} and {Ne$-$Ne} collisions at
  {$\sqrtsNN=5.36\TeV$}'',} 2025.
  \href{http://www.arXiv.org/abs/2509.06428}{\texttt{arXiv:2509.06428}}.
  Submitted to \textit{Phys. Rev. Lett.}

\bibitem{ATLAS:2025nnt}
\hrefCMSnoop {}{{ATLAS Collaboration}, ``Measurement of the azimuthal
  anisotropy of charged particles in {$\sqrtsNN=5.36\TeV$} {$^{16}$O$+^{16}$O}
  and {$^{20}$Ne$+^{20}$Ne} collisions with the {ATLAS} detector'',} 2025.
  \href{http://www.arXiv.org/abs/2509.05171}{\texttt{arXiv:2509.05171}}.
  Submitted to \textit{Phys. Rev. C}.

\bibitem{CMS:2014pgm}
\hrefCMSnoop {}{{CMS Collaboration}, ``Description and performance of track and
  primary-vertex reconstruction with the {CMS} tracker'',} \textit{ JINST}
  \textbf{ 9} (2014) P10009,
  \href{http://dx.doi.org/10.1088/1748-0221/9/10/P10009}{\doi{10.1088/1748-0221/9/10/P10009}},
  \href{http://www.arXiv.org/abs/1405.6569}{\texttt{arXiv:1405.6569}}.

\bibitem{Chatrchyan:2008zzk}
\hrefCMSnoop {}{{CMS Collaboration}, ``The {CMS} experiment at the {CERN
  LHC}'',} \textit{ JINST} \textbf{ 3} (2008) S08004,
\href{http://dx.doi.org/10.1088/1748-0221/3/08/S08004}{\doi{10.1088/1748-0221/3/08/S08004}}.

\bibitem{CMS:2023gfb}
\hrefCMSnoop {}{{CMS Collaboration}, ``Development of the {CMS} detector for
  the {CERN LHC Run} 3'',} \textit{ JINST} \textbf{ 19} (2024) P05064,
  \href{http://dx.doi.org/10.1088/1748-0221/19/05/P05064}{\doi{10.1088/1748-0221/19/05/P05064}},
  \href{http://www.arXiv.org/abs/2309.05466}{\texttt{arXiv:2309.05466}}.

\bibitem{CMS:2011iwn}
\hrefCMSnoop {}{{CMS Collaboration}, ``Observation and studies of jet quenching
  in {PbPb} collisions at nucleon-nucleon center-of-mass energy ${\sqrtsNN} =
  2.76${\TeV}'',} \textit{ Phys. Rev. C} \textbf{ 84} (2011) 024906,
  \href{http://dx.doi.org/10.1103/PhysRevC.84.024906}{\doi{10.1103/PhysRevC.84.024906}},
  \href{http://www.arXiv.org/abs/1102.1957}{\texttt{arXiv:1102.1957}}.

\bibitem{Miller:2007ri}
\hrefCMSnoop {}{M.~L. Miller, K.~Reygers, S.~J. Sanders, and P.~Steinberg,
  ``{Glauber modeling in high energy nuclear collisions}'',} \textit{ Ann. Rev.
  Nucl. Part. Sci.} \textbf{ 57} (2007) 205,
  \href{http://dx.doi.org/10.1146/annurev.nucl.57.090506.123020}{\doi{10.1146/annurev.nucl.57.090506.123020}},
  \href{http://www.arXiv.org/abs/nucl-ex/0701025}{\texttt{arXiv:nucl-ex/0701025}}.

\bibitem{Loizides:2025ule}
\hrefCMSnoop {}{C.~Loizides, ``Glauber predictions for oxygen and neon
  collisions at {LHC}'',} 2025.
  \href{http://www.arXiv.org/abs/2507.05853}{\texttt{arXiv:2507.05853}}.
  Submitted to \textit{Phys. Rev. C}.

\bibitem{CMS:2013jlh}
\hrefCMSnoop {}{{CMS Collaboration}, ``Multiplicity and transverse momentum
  dependence of two- and four-particle correlations in {\pPb} and {\PbPb}
  collisions'',} \textit{ Phys. Lett. B} \textbf{ 724} (2013) 213,
  \href{http://dx.doi.org/10.1016/j.physletb.2013.06.028}{\doi{10.1016/j.physletb.2013.06.028}},
  \href{http://www.arXiv.org/abs/1305.0609}{\texttt{arXiv:1305.0609}}.

\bibitem{Khachatryan:2014jra}
\hrefCMSnoop {}{{CMS Collaboration}, ``Long-range two-particle correlations of
  strange hadrons with charged particles in {\pPb} and {\PbPb} collisions at
  {LHC} energies'',} \textit{ Phys. Lett. B} \textbf{ 742} (2015) 200,
  \href{http://dx.doi.org/10.1016/j.physletb.2015.01.034}{\doi{10.1016/j.physletb.2015.01.034}},
\href{http://www.arXiv.org/abs/1409.3392}{\texttt{arXiv:1409.3392}}.

\bibitem{Wang:1991hta}
\hrefCMSnoop {}{X.-N. Wang and M.~Gyulassy, ``{HIJING}: A {Monte Carlo} model
  for multiple jet production in {\pp}, {\pA} and {\AonA} collisions'',}
  \textit{ Phys. Rev. D} \textbf{ 44} (1991) 3501,
  \href{http://dx.doi.org/10.1103/PhysRevD.44.3501}{\doi{10.1103/PhysRevD.44.3501}}.

\bibitem{Agostinelli:2002hh}
\hrefCMSnoop {}{{GEANT4} Collaboration, ``{\GEANTfour}---a simulation
  toolkit'',} \textit{ Nucl. Instrum. Meth. A} \textbf{ 506} (2003) 250,
\href{http://dx.doi.org/10.1016/S0168-9002(03)01368-8}{\doi{10.1016/S0168-9002(03)01368-8}}.

\bibitem{Bilandzic:2010jr}
\hrefCMSnoop {}{A.~Bilandzic, R.~Snellings, and S.~Voloshin, ``Flow analysis
  with cumulants: Direct calculations'',} \textit{ Phys. Rev. C} \textbf{ 83}
  (2011) 044913,
  \href{http://dx.doi.org/10.1103/PhysRevC.83.044913}{\doi{10.1103/PhysRevC.83.044913}},
\href{http://www.arXiv.org/abs/1010.0233}{\texttt{arXiv:1010.0233}}.

\bibitem{Lin:2004en}
Z.-W. Lin\hrefCMSnoop {}{ { et~al.}, ``A multi-phase transport model for
  relativistic heavy ion collisions'',} \textit{ Phys. Rev. C} \textbf{ 72}
  (2005) 064901,
  \href{http://dx.doi.org/10.1103/PhysRevC.72.064901}{\doi{10.1103/PhysRevC.72.064901}},
  \href{http://www.arXiv.org/abs/nucl-th/0411110}{\texttt{arXiv:nucl-th/0411110}}.

\bibitem{Pshenichnov:1997un}
I.~A. Pshenichnov\hrefCMSnoop {}{ { et~al.}, ``Nuclear multifragmentation
  induced by electromagnetic fields of ultrarelativistic heavy ions'',}
  \textit{ Phys. Rev. C} \textbf{ 57} (1998) 1920,
  \href{http://dx.doi.org/10.1103/PhysRevC.57.1920}{\doi{10.1103/PhysRevC.57.1920}},
  \href{http://www.arXiv.org/abs/nucl-th/9711030}{\texttt{arXiv:nucl-th/9711030}}.

\bibitem{Das:2025rqb}
\hrefCMSnoop {}{S.~J. Das and A.~Baty, ``A data-driven method to estimate
  contamination from light ion beam transmutation at colliders'',} 2025.
  \href{http://www.arXiv.org/abs/2509.09736}{\texttt{arXiv:2509.09736}}.

\bibitem{CMS:2013wjq}
\hrefCMSnoop {}{{CMS Collaboration}, ``Measurement of higher-order harmonic
  azimuthal anisotropy in {\PbPb} collisions at ${\sqrtsNN}$ = 2.76{\TeV}'',}
  \textit{ Phys. Rev. C} \textbf{ 89} (2014) 044906,
  \href{http://dx.doi.org/10.1103/PhysRevC.89.044906}{\doi{10.1103/PhysRevC.89.044906}},
  \href{http://www.arXiv.org/abs/1310.8651}{\texttt{arXiv:1310.8651}}.

\bibitem{Ollitrault:2009ie}
\hrefCMSnoop {}{J.-Y. Ollitrault, A.~M. Poskanzer, and S.~A. Voloshin, ``Effect
  of flow fluctuations and nonflow on elliptic flow methods'',} \textit{ Phys.
  Rev. C} \textbf{ 80} (2009) 014904,
  \href{http://dx.doi.org/10.1103/PhysRevC.80.014904}{\doi{10.1103/PhysRevC.80.014904}},
  \href{http://www.arXiv.org/abs/0904.2315}{\texttt{arXiv:0904.2315}}.

\bibitem{Nijs:2020roc}
\hrefCMSnoop {}{G.~Nijs, W.~van~der Schee, U.~G{\"u}rsoy, and R.~Snellings,
  ``Bayesian analysis of heavy ion collisions with the heavy ion computational
  framework {Trajectum}'',} \textit{ Phys. Rev. C} \textbf{ 103} (2021) 054909,
  \href{http://dx.doi.org/10.1103/PhysRevC.103.054909}{\doi{10.1103/PhysRevC.103.054909}},
  \href{http://www.arXiv.org/abs/2010.15134}{\texttt{arXiv:2010.15134}}.

\bibitem{Nijs:2021clz}
\hrefCMSnoop {}{G.~Nijs and W.~van~der Schee, ``Predictions and postdictions
  for relativistic lead and oxygen collisions with the computational simulation
  code {Trajectum}'',} \textit{ Phys. Rev. C} \textbf{ 106} (2022) 044903,
  \href{http://dx.doi.org/10.1103/PhysRevC.106.044903}{\doi{10.1103/PhysRevC.106.044903}},
  \href{http://www.arXiv.org/abs/2110.13153}{\texttt{arXiv:2110.13153}}.

\bibitem{Lahde:2019npb}
T.~A. L{\"a}hde and U.-G. Mei{\ss}ner, ``Nuclear Lattice Effective Field
  Theory: An introduction'', volume 957.
\newblock Springer, 2019.
\newblock
  \href{http://dx.doi.org/10.1007/978-3-030-14189-9}{\doi{10.1007/978-3-030-14189-9}},
  ISBN~978-3-030-14187-5, 978-3-030-14189-9.

\bibitem{Lee:2020meg}
\hrefCMSnoop {}{D.~Lee, ``Recent progress in nuclear lattice simulations'',}
  \textit{ Front. Phys.} \textbf{ 8} (2020) 174,
  \href{http://dx.doi.org/10.3389/fphy.2020.00174}{\doi{10.3389/fphy.2020.00174}}.

\bibitem{Frosini:2021tuj}
M.~Frosini\hrefCMSnoop {}{ { et~al.}, ``In-medium {$k$}-body reduction of
  {$n$}-body operators: A flexible symmetry-conserving approach based on the
  sole one-body density matrix'',} \textit{ Eur. Phys. J. A} \textbf{ 57}
  (2021) 151,
  \href{http://dx.doi.org/10.1140/epja/s10050-021-00458-z}{\doi{10.1140/epja/s10050-021-00458-z}},
  \href{http://www.arXiv.org/abs/2102.10120}{\texttt{arXiv:2102.10120}}.

\bibitem{Yao:2019rck}
J.~M. Yao\hrefCMSnoop {}{ { et~al.}, ``Ab initio treatment of collective
  correlations and the neutrinoless double beta decay of {$^{48}$Ca}'',}
  \textit{ Phys. Rev. Lett.} \textbf{ 124} (2020) 232501,
  \href{http://dx.doi.org/10.1103/PhysRevLett.124.232501}{\doi{10.1103/PhysRevLett.124.232501}},
  \href{http://www.arXiv.org/abs/1908.05424}{\texttt{arXiv:1908.05424}}.

\bibitem{Moreland:2014oya}
\hrefCMSnoop {}{J.~S. Moreland, J.~E. Bernhard, and S.~A. Bass, ``Alternative
  ansatz to wounded nucleon and binary collision scaling in high-energy nuclear
  collisions'',} \textit{ Phys. Rev. C} \textbf{ 92} (2015) 011901,
  \href{http://dx.doi.org/10.1103/PhysRevC.92.011901}{\doi{10.1103/PhysRevC.92.011901}},
  \href{http://www.arXiv.org/abs/1412.4708}{\texttt{arXiv:1412.4708}}.

\bibitem{Mantysaari:2025tcg}
\hrefCMSnoop {}{H.~M{\"a}ntysaari, B.~Schenke, C.~Shen, and W.~Zhao,
  ``Collision-energy dependence in heavy-ion collisions from nonlinear {QCD}
  evolution'',} \textit{ Phys. Rev. Lett.} \textbf{ 135} (2025) 022302,
  \href{http://dx.doi.org/10.1103/gf4y-p5j7}{\doi{10.1103/gf4y-p5j7}},
  \href{http://www.arXiv.org/abs/2502.05138}{\texttt{arXiv:2502.05138}}.

\bibitem{Schenke:2010nt}
\hrefCMSnoop {}{B.~Schenke, S.~Jeon, and C.~Gale, ``{(3+1)D} hydrodynamic
  simulation of relativistic heavy-ion collisions'',} \textit{ Phys. Rev. C}
  \textbf{ 82} (2010) 014903,
  \href{http://dx.doi.org/10.1103/PhysRevC.82.014903}{\doi{10.1103/PhysRevC.82.014903}},
  \href{http://www.arXiv.org/abs/1004.1408}{\texttt{arXiv:1004.1408}}.

\bibitem{Bass:1998ca}
\hrefCMSnoop {}{S.~A. Bass { et~al.}, ``Microscopic models for
  ultrarelativistic heavy ion collisions'',} \textit{ Prog. Part. Nucl. Phys.}
  \textbf{ 41} (1998) 255,
  \href{http://dx.doi.org/10.1016/S0146-6410(98)00058-1}{\doi{10.1016/S0146-6410(98)00058-1}},
  \href{http://www.arXiv.org/abs/nucl-th/9803035}{\texttt{arXiv:nucl-th/9803035}}.

\end{thebibliography}\endgroup

\cleardoublepage \appendix\section{The CMS Collaboration \label{app:collab}}\begin{sloppypar}\hyphenpenalty=5000\widowpenalty=500\clubpenalty=5000
\cmsinstitute{Yerevan Physics Institute, Yerevan, Armenia}
{\tolerance=6000
A.~Hayrapetyan, V.~Makarenko\cmsorcid{0000-0002-8406-8605}, A.~Tumasyan\cmsAuthorMark{1}\cmsorcid{0009-0000-0684-6742}
\par}
\cmsinstitute{Institut f\"{u}r Hochenergiephysik, Vienna, Austria}
{\tolerance=6000
W.~Adam\cmsorcid{0000-0001-9099-4341}, L.~Benato\cmsorcid{0000-0001-5135-7489}, T.~Bergauer\cmsorcid{0000-0002-5786-0293}, M.~Dragicevic\cmsorcid{0000-0003-1967-6783}, C.~Giordano, P.S.~Hussain\cmsorcid{0000-0002-4825-5278}, M.~Jeitler\cmsAuthorMark{2}\cmsorcid{0000-0002-5141-9560}, N.~Krammer\cmsorcid{0000-0002-0548-0985}, A.~Li\cmsorcid{0000-0002-4547-116X}, D.~Liko\cmsorcid{0000-0002-3380-473X}, M.~Matthewman, J.~Schieck\cmsAuthorMark{2}\cmsorcid{0000-0002-1058-8093}, R.~Sch\"{o}fbeck\cmsAuthorMark{2}\cmsorcid{0000-0002-2332-8784}, M.~Shooshtari\cmsorcid{0009-0004-8882-4887}, M.~Sonawane\cmsorcid{0000-0003-0510-7010}, W.~Waltenberger\cmsorcid{0000-0002-6215-7228}, C.-E.~Wulz\cmsAuthorMark{2}\cmsorcid{0000-0001-9226-5812}
% ...
\par}
\vskip\cmsinstskip
\dag:~Deceased\\
$^{1}$Also at Yerevan State University, Yerevan, Armenia\\
$^{2}$Also at TU Wien, Vienna, Austria\\
% ...
\end{sloppypar}
\end{document}